\documentstyle[11pt]{article}
\newcommand{\be}{\begin{eqnarray}}
\newcommand{\ee}{\end{eqnarray}}
\newcommand\non {\nonumber}
\newcommand\uh {\frac u 2}

\newcommand\half{\frac 1 2 }
\newcommand\noi{\noindent}
\newcommand\tr{\mathop{\rm tr}\nolimits}
\font\zch=PSMZCHMI scaled 1000
\newcommand\g{\hbox{\zch g}}

\begin{document}

\begin{titlepage}
\strut\hfill BONN--TH--94--19, UMTG--178
\vspace{.5in}
\begin{center}
\LARGE Spectrum of transfer matrix for $U_q(B_n)$-invariant \\
\LARGE $A^{(2)}_{2n}$ open spin chain\\[.2in]
\large Simone Artz\footnote{Physics Department, Bonn University,
      Nussallee 12, D-53115 Bonn, Germany}\hfill
 Luca Mezincescu${}^2$\hfill
 Rafael I. Nepomechie\footnote{Physics Department, University of Miami,
      Coral Gables, FL 33124 USA}
\end{center}
\vspace{.5in}
\begin{abstract}
We propose an expression for the eigenvalues of the transfer matrix for
the $U_q(B_n)$-invariant open quantum spin chain associated with the
fundamental representation of $A^{(2)}_{2n}$. By assumption, the Bethe
Ansatz equations are ``doubled'' with respect to those of the corresponding
closed chain with periodic boundary conditions. We verify that
the transfer matrix eigenvalues have the correct analyticity properties
and asymptotic behavior. We also briefly discuss the structure of
the eigenstates of the transfer matrix.
\end{abstract}
\end{titlepage}

\setcounter{footnote}{0}
\section{Introduction}

Integrable open quantum spin chains with quantum algebra \cite{qsymmetry}
symmetry provide interesting examples of quantum-algebra-invariant physical
systems. Indeed, these models have rich structure (nontrivial vacuum,
excitations with nontrivial scattering, etc.) which can nevertheless be
investigated in detail by virtue of the models' integrability.
Such models have therefore been studied intensively.
(See, e.g., Refs. \cite{alcaraz} - \cite{karowski/zapletal}.)
In particular, the eigenvalues of the transfer matrix (in terms of solutions
of the corresponding Bethe Ansatz equations) have been obtained for the
following models:
\begin{itemize}
\item $U_q(A_1)$-invariant $A^{(1)}_{1}$ chain with spin 1/2
\cite{sklyanin} and with spin 1\cite{spin1} (see also
Refs. \cite{kulish/sklyanin(1),mpla,destri/devega})
\item $U_q(A_1)$-invariant $A^{(2)}_{2}$ chain \cite{I}
\item $U_q(A_n)$-invariant $A^{(1)}_{n}$ chain
\cite{devega/ruiz(1),devega/ruiz(2)}
\item $U_q(spl(2,1))$-invariant t-J model \cite{foerster/karowski,ruiz}
\end{itemize}
For all these cases, the Bethe ansatz equations
are ``doubled'' with respect to those of the corresponding closed
chains with periodic boundary conditions.

In this paper, we study the $U_q(B_n)$-invariant integrable
open quantum spin chain which is constructed from the $R$ matrix
associated with the fundamental representation of $A^{(2)}_{2n}$.
In particular, we propose an expression for the exact eigenvalues of the
transfer matrix for the finite chain. This generalizes the result for the
case $n=1$ given in Ref. \cite{I}.
We also briefly discuss the structure of the eigenstates of
the transfer matrix.

Our approach is a variation of the analytical Bethe Ansatz method,
whereby the transfer matrix eigenvalues are assumed to have the form
of ``dressed'' pseudovacuum eigenvalues; and the dressing functions are
determined from the requirements of analyticity, asymptotic behavior,
crossing, etc.
This method was developed in Refs. \cite{analytical,reshetikhin}
for integrable {\em closed} chains,
and then generalized in \cite{I} for integrable quantum-algebra invariant
{\em open} chains. For the case $A^{(2)}_{2n}$ with $n>1$, a
direct calculation of the pseudovacuum eigenvalue appears to be difficult.
We proceed to determine the transfer matrix eigenvalues instead
by demanding that the Bethe Ansatz equations be ``doubled'' with
respect to those of the closed chain. (As already
noted, this is a characteristic feature of all the previously
studied cases.) This procedure uniquely fixes the eigenvalues of
the transfer matrix. We verify that our result is consistent
with the requirements of analyticity, asymptotic behavior, etc.
We expect that this procedure can be applied to all the
quantum-algebra invariant integrable open chains constructed from
trigonometric $R$ matrices.

The paper is organized as follows. In Section 2, we define the
model and formulate the problem. In Section 3, we arrive at
the expression for the eigenvalues of the transfer matrix.
We perform a number of checks on this result in Section 4.
We briefly discuss some remaining questions in Section 5. There
are two appendices. In Appendix A we summarize the properties of
the $A^{(2)}_{2n}$ $R$ matrix. In Appendix B, we briefly
discuss the structure of the eigenstates of the transfer matrix.

\section{The problem}

We consider the open quantum spin chain with
Hamiltonian \cite{ijmpa,mpla,I}
\be
{\cal H} = \sum_{k=1}^{N-1} {d\over du} \check R_{k,k+1}(u) \Big\vert_{u=0}
\,. \label{hamiltonian}
\ee
We follow the standard notations (see, e.g., \cite{kulish/sklyanin(2)}).
In particular,
$\check R(u) = {\cal P} R(u)$, where ${\cal P}$ is the permutation matrix.
Furthermore, $R(u)$ is an $R$ matrix; i.e., it obeys the Yang-Baxter equation
\be
R_{12}(u - v)\  R_{13}(u)\ R_{23}(v) = R_{23}(v)\  R_{13}(u)\ R_{12}(u - v)
\,.   \label{yang-baxter}
\ee

We restrict our attention here to the case that $R(u)$ is the
$R$ matrix associated with the fundamental
representation of $A^{(2)}_{2n}$. This $R$ matrix, which is a
generalization of the $A^{(2)}_{2}$ $R$ matrix of Izergin and Korepin
\cite{izergin/korepin}, was found by Bazhanov \cite{bazhanov}
and Jimbo \cite{jimbo}. It depends on an anisotropy parameter $\eta$,
which we take to be real. In Appendix A, we reproduce this
$R$ matrix and summarize its important properties. In the present
section, it suffices to point out that the $R$ matrix is
crossing symmetric
\be
R_{12}(u) = V_1 \ R_{12}(-u - \rho)^{t_2}\  V_1
=  V_2^{t_2} \ R_{12}(-u - \rho)^{t_1}\  V_2^{t_2}  \,,
\label{crossing}
\ee
where the crossing parameter $\rho$ is given by
\be
\rho=-i \pi -2(2n+1)\eta \,,
\ee
$t_i$ denotes transposition in the $i^{th}$ space, and the crossing
matrix $V$ satisfies
$V^2 = 1$ (we use the notation $V_1 = V \otimes 1$, $V_2 = 1 \otimes V$).

The Hamiltonian has \cite{jimbo,ijmpa} the quantum-algebra symmetry $U_q(B_n)$
\be
\left[ {\cal H} \,, U_q (B_n) \right] = 0 \,, \label{qinvariance0}
\ee
$B_n$ (= $O(2n+1)$) being the maximal finite-dimensional subalgebra of the
affine algebra $A^{(2)}_{2n}$. More precisely, ${\cal H}$ commutes
with $\Delta^{(N)}(x)$, the $N$-fold coproduct of any generator $x$
of $U_q (B_n)$ in the fundamental ($(2n+1)$-dimensional) representation.
The real parameter $q$ is related to the anisotropy parameter
$\eta$ by $q=e^{2\eta}$.
Evidently, the Hamiltonian (\ref{hamiltonian}) describes a set of $N$ spins
in the fundamental representation of $U_q(B_n)$, with nearest-neighbor
interactions.
The state space of the Hamiltonian is the $N$-fold tensor product space
$C^{2n+1}\otimes \cdots \otimes C^{2n+1}$.

This open spin chain is integrable \cite{ijmpa}. The
transfer matrix is given by \cite{sklyanin,ijmpa,I}
\footnote{By working in the so-called homogeneous gauge, the $R$
matrix has the commutativity property \cite{jimbo}
$\left[ \check R(u) \,, \check R(v) \right] = 0$. This implies
\cite{ijmpa} that the equations \cite{sklyanin,jpa} for the reflection
matrix (boundary $S$ matrix) $K_-(u)$ have the solution $K_-(u)=1$, which
is the one corresponding to the Hamiltonian (\ref{hamiltonian}).}
\be
t(u) = \tr_a M_a\ T_a(u)\  \hat T_a(u) \,, \label{transfer}
\ee
where
\begin{eqnarray}
     T_a(u) &=& R_{aN}(u)\ R_{a N-1}(u)\ \cdots R_{a1}(u)
\,,  \non \\
\hat T_a(u) &=& R_{1a}(u)\ \cdots R_{N-1 a}(u)\ R_{Na}(u)
\,,  \label{monodromy}
\end{eqnarray}
and the matrix $M$ is given by
\be
M = V^t\ V = M^t \,,
\ee
where $V$ is the crossing matrix introduced in Eq. (\ref{crossing}).
The subscript $a$ denotes the auxiliary space. (As usual, we suppress the
quantum-space subscripts $1 \,, \cdots \,, N$ of $T_a(u)$ and $\hat T_a(u)$.)
The transfer matrix is also equal to \cite{I}
\be
t(u) = \tr_a M_a^{-1}\ \hat T_a(u)\ T_a(u)
\,. \label{transfertoo}
\ee

One can show \cite{sklyanin,ijmpa} that
$t(u)$ constitutes a one-parameter commutative family
\be
\left[ t(u) \,, t(v) \right] = 0 \hbox{   for all   } u \,, v
\,, \label{commutativity}
\ee
which is related to the Hamiltonian (\ref{hamiltonian}) by
\be
{\cal H} \propto {d\over du} t(u) \Big\vert_{u=0} \,.
\ee

Not only the Hamiltonian but also the transfer matrix
is $U_q(B_n)$ invariant \cite{kulish/sklyanin(1),mpla}
\be
\left[ t(u) \,, U_q (B_n) \right] = 0 \,. \label{qinvariance}
\ee
Evidently, this relation implies Eq. (\ref{qinvariance0}).

Let $\left\{ H_l \right\} \,, \quad l = 1 \,, \cdots \,, n$
be the (Hermitian) generators of the Cartan subalgebra of $U_q(B_n)$.
That is,
\be
H_l = \sum_{k=1}^N H_{l (k)} \,, \qquad l = 1\,, \cdots \,, n\,,
\label{cartan}
\ee
where $\left\{ H_{l (k)} \right\} \,, \quad l = 1 \,, \cdots \,, n$
are the corresponding generators in the fundamental representation
at site $k$. In the standard basis, these generators are given by
the $(2n+1) \times (2n+1)$ diagonal matrices
\be
H_{1 (k)}   &=& diag \left( 1 \,, 0 \,, \cdots \,, 0 \,, -1 \right)_{k}
\,, \non  \\
             & \vdots & \label{hl} \\
H_{n-1 (k)} &=& diag \left( 0 \,, \cdots \,, 0 \,, 1 \,, 0 \,,
0 \,, 0 \,, -1 \,, 0\,, \cdots \,, 0 \right)_{k} \,, \non \\
H_{n (k)}   &=& diag \left( 0 \,, \cdots \,, 0 \,, 1 \,, 0 \,, -1 \,,
0\,, \cdots \,, 0 \right)_{k} \,.  \non
\ee

Since the $R$ matrix is real, the transfer matrix $t(u)$ is Hermitian.
It follows that there exist simultaneous eigenstates
$| \Lambda^{(m_1 \,, \cdots \,, m_n)} >$ of $t(u)$ and $\left\{ H_l \right\}$:
\begin{eqnarray}
t(u)\ | \Lambda^{(m_1 \,, \cdots \,, m_n)} > &=&
\Lambda^{(m_1 \,, \cdots \,, m_n)}(u)\
| \Lambda^{(m_1 \,, \cdots \,, m_n)} > \,, \non \\
 H_l\ | \Lambda^{(m_1 \,, \cdots \,, m_n)} > &=&  \lambda_l\
| \Lambda^{(m_1 \,, \cdots \,, m_n)} > \,, \quad\quad l = 1 \,, \cdots \,, n
\,.
\end{eqnarray}
We choose these states to be highest weights of $U_q(B_n)$.
(See e.g. Refs. \cite{mpla,I,destri/devega,devega/ruiz(2)}.)
The eigenvalues $\lambda_l$  are related to the nonnegative
integers $m_1 \,, \cdots \,, m_n$ by
\begin{eqnarray}
\lambda_1 &=& N - m_1 \,, \non     \\
\lambda_2 &=& m_1 - m_2 \,,    \non          \\
          &\vdots &             \label{r}     \\
\lambda_n &=& m_{n-1} - m_n \,, \non
\end{eqnarray}
These relations, which are the same as for the closed
$A^{(2)}_{2n}$ chain \cite{reshetikhin}, are explained
in Appendix B.

Our problem is to determine $\Lambda^{(m_1 \,, \cdots \,, m_n)}(u)$,
the spectrum of the transfer matrix $t(u)$. In the next Section, we
propose a solution to this problem.

\section{The proposed solution}

Our approach to finding the eigenvalues of the transfer matrix
is a variation of the analytical Bethe Ansatz method. We
shall need the following two results: the crossing relation \cite{I}
\be
\Lambda^{(m_1 \,, \cdots \,, m_n)}(u) =
   \Lambda^{(m_1 \,, \cdots \,, m_n)}(-u - \rho) \,,
\label{crossing-lambda}
\ee
and the fusion formula \cite{fusion,I}
\be
\tilde \Lambda^{(m_1 \,, \cdots \,, m_n)}(u) & = &
{1\over \alpha(u)^{2N}\ \beta(u)^2}
\big\{ \zeta( 2u + 2\rho)\ \Lambda^{(m_1 \,, \cdots \,, m_n)}(u)\
\Lambda^{(m_1 \,, \cdots \,, m_n)}(u+\rho) \non\\
& & - \zeta( u + \rho)^{2N}\  \g(2u + \rho)\ \g(-2u - 3\rho) \big\}
\,. \label{fusion}
\ee
$\tilde \Lambda^{(m_1 \,, \cdots \,, m_n)}(u)$ is the eigenvalue
of the so-called fused transfer matrix, which is obtained by
fusion \cite{kulish/sklyanin(2),ffusion}
in the auxiliary space. Both $\Lambda^{(m_1 \,, \cdots \,, m_n)}(u)$
and $\tilde \Lambda^{(m_1 \,, \cdots \,, m_n)}(u)$ are analytic
functions of $u$. (See, e.g., Refs. \cite{analytical,I}.)
The function $\zeta(u)$ is given by
\be
\zeta(u) = \g(u)\ \g(-u) \,,
\ee
and the function $\g(u)$ is defined by
\be
\g(u) & = & \tr_{12}\ R_{12}(u)\ V_1\ V_2\ \tilde P_{12}^- \,, \non \\
      & = & 2 \sinh (\uh -2\eta)\ \cosh(\uh + (2n+1)\eta)  \,. \label{g}
\ee
The projector $\tilde P_{12}^-$ is given by
\be
\tilde P_{12}^- \propto R_{12}(-\rho) \,,
\ee
where the normalization constant is determined by the
condition $\left( \tilde P_{12}^- \right)^2 = \tilde P_{12}^-$.
Finally, $\alpha(u)$ and $\beta(u)$ are functions that vanish
at $u = -\rho$ which are given by
\be
\alpha(u)&=&\cosh(\uh-(2n+1)\eta) \,, \non \\
\beta(u) &=&\sinh(u-2(2n+1)\eta)  \,.
\ee

Consider now the so-called pseudovacuum state
$| \Lambda^{(0 \,, \cdots \,, 0)} >$, which
is the state with all spins up,
\be
| \Lambda^{(0 \,, \cdots \,, 0)}> =
\left( \begin{array}{l}
          1 \\
          0 \\
          \vdots \\
          0
        \end{array} \right) \otimes \cdots \otimes
\left( \begin{array}{l}
           1 \\
           0 \\
           \vdots \\
           0
        \end{array} \right)
\,. \label{up}
\ee
As remarked in the Introduction, a direct calculation of the
matrix element
\be
\Lambda^{(0 \,, \cdots \,, 0)}(u)&=&
<\Lambda^{(0 \,, \cdots \,, 0)}|t(u)|\Lambda^{(0 \,, \cdots \,, 0)}>
\non \\
&=&\tr_a\ M_a <\Lambda^{(0 \,, \cdots \,, 0)}|T_a(u)\ \hat{T}_a(u)
|\Lambda^{(0 \,, \cdots \,, 0)}>
\label{matrixelement}
\ee
appears to be difficult for $n>1$. Nevertheless, it is easy to guess a
part of it. Indeed, inspection of the known result for
$n=1$ \cite{I} suggests that the pseudovacuum
eigenvalue (for general $n$) is given by
\be
\Lambda^{(0 \,, \cdots \,, 0)}(u) = f(u) + f(-u - \rho)
+ r(u)\ b(u)^{2N} \,,
\label{pseudoeigenvalue}
\ee
where $f(u)$ is given by
\be
f(u) = {\g (-2u - \rho) \over c(2u)} c(u)^{2N} \,, \label{f}
\ee
$\g(u)$ is given in Eq. (\ref{g}), and $c(u)$ is given by (see
Eq. (\ref{factors}) )
\be
c(u) = 2 \sinh (\uh -2\eta)\ \cosh(\uh - (2n+1)\eta)  \,. \label{c}
\ee
We remark that
\be
c(u)\ c(-u) = \zeta(u) \,.
\ee
Moreover, $b(u)$ is given by
\be
b(u) = 2 \sinh (\uh)\ \cosh(\uh - (2n+1)\eta)  \,. \label{b}
\ee
The crossing relation (\ref{crossing-lambda}) implies that the
function $r(u)$ is invariant under crossing
($r(u) = r(-u - \rho)$), but it is otherwise not yet specified.

The expression (\ref{pseudoeigenvalue}) for the pseudovacuum eigenvalue
is consistent with the fusion formula (\ref{fusion}). Indeed, on the RHS
of Eq. (\ref{fusion}), the last part cancels with a term occurring in
the first part. This cancelation is a necessary condition for
$\tilde \Lambda^{(m_1 \,, \cdots \,, m_n)}(u)$ to be analytic
at $u = -\rho$.

Following the analytical Bethe Ansatz approach, we assume that a
general eigenvalue $\Lambda^{(m_1 \,, \cdots \,, m_n)}(u)$ has the form
of a ``dressed'' pseudovacuum eigenvalue. That is, we make the ansatz
\be
& & \Lambda^{(m_1 \,, \cdots \,, m_n)}(u) \non \\
&=& A^{(m_1)}(u)\
\frac{\sinh(u-2(2n+1)\eta)}{\sinh(u-2\eta)}
\frac{\cosh(u-(2n-1)\eta)}{\cosh(u-(2n+1)\eta)} \non \\
& &\qquad \times
\left[ 2 \sinh (\uh -2\eta) \cosh(\uh - (2n+1)\eta) \right]^{2N} \non \\
&+&C^{(m_1)}(u)\
\frac{\sinh u}{\sinh (u-4n\eta)}
\frac{\cosh(u-(2n+3)\eta)}{\cosh(u-(2n+1)\eta)}  \non \\
& &\qquad \times
\left[ 2 \sinh (\uh) \cosh(\uh - (2n-1)\eta)\right]^{2N} \non \\
&+& \left\{ w(u)\ B_n^{(m_n)}(u) +
\sum_{l=1}^{n-1} \left[ z_l(u)\ B_l^{(m_l \,, m_{l+1})}(u)
+ \tilde z_l(u)\ \tilde B_l^{(m_l \,, m_{l+1})}(u) \right] \right\}
\non\\
& &\qquad \times \left[2 \sinh(\uh) \cosh(\uh -(2n+1)\eta)\right]^{2N}
\,.
\label{ansatz}
\ee

In contrast to the usual situation \cite{analytical,reshetikhin,I},
the ansatz involves also a set of unknown functions which are
{\em independent} of $m_1 \,, \cdots \,, m_n$ (namely, $w(u)$ and $z_l(u) \,,
\tilde z_l(u) \,, \quad l = 1 \,, \cdots \,, n-1$). This is a reflection
of the fact that the function $r(u)$ in the pseudovacuum eigenvalue
(\ref{pseudoeigenvalue}) is not known.

We shall determine all the unknown functions by the requirement that
the Bethe Ansatz equations be ``doubled'' with respect to those of
the closed $A^{(2)}_{2n}$ chain with periodic boundary conditions.
(By ``doubled'', we mean that $N$ is replaced by $2N$; and for every
factor involving the difference of rapidities, there is a similar factor
involving the {\em sum} of rapidities.)
As noted in the Introduction, this is a characteristic feature of all
the previously studied cases. By so ``doubling'' the Bethe Ansatz equations
of Reshetikhin \cite{reshetikhin}, we obtain (apart from some minor changes
in notation) the following system of Bethe Ansatz equations:
\be
\left[
\frac{\sinh({u_k^{(1)}\over 2}-\eta)}
     {\sinh({u_k^{(1)}\over 2}+\eta)}\right]^{2N}
&=&\prod_{j \ne k}^{m_1}
\frac{\sinh({1\over 2}({u_k^{(1)}-u_j^{(1)}})-2\eta)\
      \sinh({1\over 2}({u_k^{(1)}+u_j^{(1)}})-2\eta)}
     {\sinh({1\over 2}({u_k^{(1)}-u_j^{(1)}})+2\eta)\
      \sinh({1\over 2}({u_k^{(1)}+u_j^{(1)}})+2\eta)} \non \\
&\times &\prod_{j=1}^{m_2}
\frac{\sinh({1\over 2}({u_k^{(1)}-u_j^{(2)}})+\eta)\
      \sinh({1\over 2}({u_k^{(1)}+u_j^{(2)}})+\eta)}
     {\sinh({1\over 2}({u_k^{(1)}-u_j^{(2)}})-\eta)\
      \sinh({1\over 2}({u_k^{(1)}+u_j^{(2)}})-\eta)} \,, \non
\ee

\be
1&=&  \prod_{j=1}^{m_{l-1}}
\frac{\sinh({1\over 2}({u_k^{(l)}-u_j^{(l-1)}})+\eta)\
      \sinh({1\over 2}({u_k^{(l)}+u_j^{(l-1)}})+\eta)}
     {\sinh({1\over 2}({u_k^{(l)}-u_j^{(l-1)}})-\eta)\
      \sinh({1\over 2}({u_k^{(l)}+u_j^{(l-1)}})-\eta)}\non\\
&\times &  \prod_{j \ne k}^{m_l}
\frac{\sinh({1\over 2}({u_k^{(l)}-u_j^{(l)}})-2\eta)\
      \sinh({1\over 2}({u_k^{(l)}+u_j^{(l)}})-2\eta)}
     {\sinh({1\over 2}({u_k^{(l)}-u_j^{(l)}})+2\eta)\
      \sinh({1\over 2}({u_k^{(l)}+u_j^{(l)}})+2\eta)} \label{ba} \\
& \times & \prod_{j=1}^{m_{l+1}}
\frac{\sinh({1\over 2}({u_k^{(l)}-u_j^{(l+1)}})+\eta)\
      \sinh({1\over 2}({u_k^{(l)}+u_j^{(l+1)}})+\eta)}
     {\sinh({1\over 2}({u_k^{(l)}-u_j^{(l+1)}})-\eta)\
      \sinh({1\over 2}({u_k^{(l)}+u_j^{(l+1)}})-\eta)} \,, \non\\
& & \qquad \qquad l=2\,, \cdots \,, n-2 \,,  \non
\ee

\be
1&=&   \prod_{j=1}^{m_{n-1}}
\frac{\sinh({1\over 2}({u_k^{(n)}-u_j^{(n-1)}})+\eta)\
      \sinh({1\over 2}({u_k^{(n)}+u_j^{(n-1)}})+\eta)}
     {\sinh({1\over 2}({u_k^{(n)}-u_j^{(n-1)}})-\eta)\
      \sinh({1\over 2}({u_k^{(n)}+u_j^{(n-1)}})-\eta)}\non\\
& \times&  \prod_{j \ne k}^{m_n}
\frac{\sinh({1\over 2}({u_k^{(n)}-u_j^{(n)}})-2\eta)\
      \sinh({1\over 2}({u_k^{(n)}+u_j^{(n)}})-2\eta)}
     {\sinh({1\over 2}({u_k^{(n)}-u_j^{(n)}})+2\eta)\
      \sinh({1\over 2}({u_k^{(n)}+u_j^{(n)}})+2\eta)} \non \\
& & \times
\frac{\cosh({1\over 2}({u_k^{(n)}-u_j^{(n)}})+\eta)\
      \cosh({1\over 2}({u_k^{(n)}+u_j^{(n)}})+\eta)}
     {\cosh({1\over 2}({u_k^{(n)}-u_j^{(n)}})-\eta)\
      \cosh({1\over 2}({u_k^{(n)}+u_j^{(n)}})-\eta)} \non \,.
\ee

\noi
Therefore, in the ansatz (\ref{ansatz}), we set
\be
A^{(m_1)}(u)&=&\prod_{j=1}^{m_1}
\frac{\sinh({1\over 2}({u-u_j^{(1)}})+\eta)\
      \sinh({1\over 2}({u+u_j^{(1)}})+\eta)}
     {\sinh({1\over 2}({u-u_j^{(1)}})-\eta)\
      \sinh({1\over 2}({u+u_j^{(1)}})-\eta)}\,, \non \\
\non\\
C^{(m_1)}(u)&=&A^{(m_1)}(-u-\rho)  \non \\
&=&\prod_{j=1}^{m_1}
\frac{\cosh({1\over 2}({u-u_j^{(1)}})-2(n+1)\eta)\
      \cosh({1\over 2}({u+u_j^{(1)}})-2(n+1)\eta)}
     {\cosh({1\over 2}({u-u_j^{(1)}})-2n\eta)\
      \cosh({1\over 2}({u+u_j^{(1)}})-2n\eta)}\,, \non \\
\non\\
B_l^{(m_l \,, m_{l+1})}(u)&=& \prod_{j=1}^{m_l}
\frac{\sinh({1\over 2}({u-u_j^{(l)}})-(l+2)\eta)\
      \sinh({1\over 2}({u+u_j^{(l)}})-(l+2)\eta)}
     {\sinh({1\over 2}({u-u_j^{(l)}})-l\eta)\
      \sinh({1\over 2}({u+u_j^{(l)}})-l\eta)} \non\\
&\times & \prod_{j=1}^{m_{l+1}}
\frac{\sinh({1\over 2}({u-u_j^{(l+1)}})-(l-1)\eta)\
      \sinh({1\over 2}({u+u_j^{(l+1)}})-(l-1)\eta)}
     {\sinh({1\over 2}({u-u_j^{(l+1)}})-(l+1)\eta)\
      \sinh({1\over 2}({u+u_j^{(l+1)}})-(l+1)\eta)} \non \\
& & \qquad \qquad \qquad \qquad l=1\,, \cdots \,, n-1 \,, \non \\
\non\\
\tilde B_l^{(m_l \,, m_{l+1})}(u)&=& B_l^{(m_l \,, m_{l+1})}(-u-\rho) \,,
 \qquad l=1\,, \cdots \,, n-1 \,,  \label{A-B}\\
\non\\
B_n^{(m_n)}(u)&=&\prod_{j=1}^{m_n}
\frac{\sinh({1\over 2}({u-u_j^{(n)}})-(n+2)\eta)\
      \sinh({1\over 2}({u+u_j^{(n)}})-(n+2)\eta)}
     {\sinh({1\over 2}({u-u_j^{(n)}})-n\eta)\
      \sinh({1\over 2}({u+u_j^{(n)}})-n\eta)}\non\\
& & \times
\frac{\cosh({1\over 2}({u-u_j^{(n)}})-(n-1)\eta)\
      \cosh({1\over 2}({u+u_j^{(n)}})-(n-1)\eta)}
     {\cosh({1\over 2}({u-u_j^{(n)}})-(n+1)\eta)\
      \cosh({1\over 2}({u+u_j^{(n)}})-(n+1)\eta)} \,. \non
\ee
These expressions are the ``doubles'' of the corresponding quantities
for the closed $A^{(2)}_{2n}$ chain given by Reshetikhin \cite{reshetikhin}
(apart from some minor changes in notation), and so are invariant
under $u_j^{(l)} \rightarrow - u_j^{(l)}$.

The remaining unknown functions in the ansatz (\ref{ansatz})
now become uniquely determined by demanding that
the Bethe Ansatz equations (\ref{ba}) be
the conditions for the residues of $\Lambda^{(m_1 \,, \cdots \,, m_n)}(u)$
to vanish at the poles $u = u_k^{(l)} + 2 l \eta \,, \quad
l = 1 \,, \cdots \,, n$. We obtain
\be
z_l(u) & = &  \frac{\sinh(u)}{\sinh(u-2l\eta)}
\frac{\sinh(u-2(2n+1)\eta)}{\sinh(u-2(l+1)\eta)}
\frac{\cosh(u-(2n-1)\eta)}{\cosh(u-(2n+1)\eta)} \,,  \non \\
\tilde z_l(u) &=& z_l(-u -\rho)                 \,,  \non \\
& & \qquad\qquad  l = 1 \,, \cdots \,, n-1\,, \label{z-w}  \\
w(u) & = & \frac{\sinh(u)}{\sinh(u-2n\eta)}
\frac{\sinh(u-2(2n+1)\eta)}{\sinh(u-2(n+1)\eta)} \,. \non
\ee

The Eqs. (\ref{ansatz}), (\ref{A-B}), and (\ref{z-w})
for the transfer matrix eigenvalues are the main result of this paper.
Since our procedure is far from rigorous, it is desirable to
perform some checks on the result. This is the subject of the next section.

\section{Checks}

In this section, we perform various checks on our expression for
the transfer matrix eigenvalues. First, it is easy to verify that
for $n=1$ our result reduces to the result given in \cite{I}.
Also, the crossing relation (\ref{crossing-lambda}) is satisfied
(in particular, $w(u) = w(-u - \rho)$). Moreover, our expression
for the transfer matrix eigenvalues is a periodic function of $u$,
with period $2\pi i$, as required by the corresponding
periodicity of the $R$ matrix.

As already remarked, the transfer matrix eigenvalues must be analytic
functions of $u$. One can verify that the conditions for the residues of
$\Lambda^{(m_1 \,, \cdots \,, m_n)}(u)$ to vanish at the poles
$u = -u_k^{(l)} + 2 l \eta \,, \quad l = 1 \,, \cdots \,, n$
\footnote{These poles, which appear due to the ``doubling'' of the
Bethe Ansatz equations, are not present for the closed chain.}
are the {\em same} system of Bethe Ansatz equations (\ref{ba}).
Moreover, the conditions for the residues of
$\Lambda^{(m_1 \,, \cdots \,, m_n)}(u)$ to vanish at the poles
$u = 2 l \eta \,, \quad l = 1 \,, \cdots \,, n$ are (see Eqs.
(\ref{ansatz}), (\ref{z-w}))
\be
A^{(m_1)}(2\eta) &=& B_1^{(m_1 \,, m_{2})}(2\eta) \,, \non \\
B_{l-1}^{(m_{l-1} \,, m_{l})}(2l\eta) &=&
B_l^{(m_l \,, m_{l+1})}(2l\eta) \,, \qquad l = 2\,, \cdots \,, n
\ee
One can verify that these conditions are satisfied. Hence, our
result for $\Lambda^{(m_1 \,, \cdots \,, m_n)}(u)$ satisfies the
requirement of analyticity.

A necessary condition for the expression (\ref{fusion}) for the
eigenvalues of the fused transfer matrix
$\tilde \Lambda^{(m_1 \,, \cdots \,, m_n)}(u)$ to be
analytic at $u = -\rho$ is
\be
A^{(m_1)}(u + \rho)\ C^{(m_1)}(u) = 1 \,.
\ee
This condition is satisfied. Moreover, we have explicitly verified
that this expression for $\tilde \Lambda^{(m_1 \,, \cdots \,, m_n)}(u)$
is in fact analytic at $u = -\rho$.

Finally, we observe that the leading asymptotic behavior of
$\Lambda^{(m_1 \,, \cdots \,, m_n)}(u)$ for large $u$ can
be computed directly, without reference to the ansatz
(\ref{ansatz}). The key point is that, for large $u$,
the monodromy matrix can be expressed in terms of generators
of $U_q(B_n)$.
Indeed, recalling the definition (\ref{monodromy})
of the monodromy matrix $T_a(u)$ and the asymptotic behavior
(\ref{Rasymptotic}) of the $R$ matrix, we obtain
\be
T_a(u) \stackrel{u \rightarrow \infty}{\sim}
\kappa(u)\ T_a^+  \,,
\ee
where $T_a^+$ is a $u$-independent, lower-triangular matrix
given by
\be
T_a^+ = \left( \begin{array}{lcccccr}
e^{-2\eta H_1} &       &         &      &        &     & \\
      &      \ddots    &         &      &        0     & \\
      &        &  e^{-2\eta H_n} &      &        &     & \\
      &        &       &         1      &        &     & \\
      &        &       &         & e^{2\eta H_n} &     & \\
      &      \star     &         &      &      \ddots  & \\
      &        &       &         &      &        & e^{2\eta H_1}
\end{array}\right) \label{Tasymptotic} \,,
\ee
whose off-diagonal elements (represented by $\star$)
involve raising operators of $U_q(B_n)$; the scalar function
$\kappa(u)$ is given by
\be
\kappa(u)=\left[ \half e^{u-(2n+1)\eta}\right]^N \,,
\ee
and $H_1 \,, \cdots \,, H_n$ are the Cartan generators of $U_q(B_n)$
given in Eq. (\ref{cartan}). Similarly,
\be
\hat T_a(u) \stackrel{u \rightarrow \infty}{\sim}
\kappa(u)\ \hat T_a^+ \,,
\ee
where $\hat T_a^+$ is the transpose (in both auxiliary and quantum spaces)
of $T_a^+$.

Since the states $| \Lambda^{(m_1 \,, \cdots \,, m_n)} >$
are highest weights of $U_q(B_n)$, the action of $T_a^+$ on a
state gives a {\em diagonal} matrix
\be
T_a^+ \ | \Lambda^{(m_1 \,, \cdots \,, m_n)} > =
\left( \begin{array}{lcccccr}
e^{-2\eta\lambda_1}&     &            &       &            &        & \\
      &      \ddots      &            &       &            0        & \\
      &        &  e^{-2\eta\lambda_n} &       &            &        & \\
      &        &         &            1       &            &        & \\
      &        &         &            & e^{2\eta\lambda_n} &        & \\
      &        0         &            &       &         \ddots      & \\
      &        &         &            &       &            & e^{2\eta\lambda_1}
\end{array}\right) | \Lambda^{(m_1 \,, \cdots \,, m_n)} > \,,
\ee
where the eigenvalues $\{ \lambda_l \}$ are related to the integers
$\{ m_l \}$ by Eq. (\ref{r}).

Making use of the second form (\ref{transfertoo}) of the transfer matrix
and the explicit expression (\ref{M}) for the matrix $M$,
we obtain the leading asymptotic behavior of
$\Lambda^{(m_1 \,, \cdots \,, m_n)}(u)$ for large $u$:
\be
\Lambda^{(m_1 \,, \cdots \,, m_n)}(u)
&\stackrel{u \rightarrow \infty}{\sim}
&\kappa(u)^2\ \tr_a M_a^{-1}
<\Lambda^{(m_1 \,, \cdots \,, m_n)}|\hat{T}_a^+\ T_a^+
|\Lambda^{(m_1 \,, \cdots \,, m_n)}> \non \\
&=&\left[ \half e^{u-(2n+1)\eta}\right]^{2N}
\big\{ 1 + \sum_{m=1}^n \big[ e^{-2(2n+1-2m)\eta - 4\eta\lambda_m} \non\\
& & + e^{2(2n+1-2m)\eta + 4\eta\lambda_m} \big] \big\} \,.
\label{asymptotic}
\ee

Precisely the same result is obtained from our expression (\ref{ansatz}),
as one can readily check using the following $u \rightarrow \infty$ limits:
\be
\begin{array}{clccclc}
A^{(m_1)}(u) &\rightarrow &e^{4\eta m_1}  &\,, \qquad
&C^{(m_1)}(u) &\rightarrow &e^{-4\eta m_1} \,, \non \\
B_l^{(m_l \,, m_{l+1})}(u) &\rightarrow &e^{-4\eta (m_l - m_{l+1})}
&\,, \qquad
&\tilde B_l^{(m_l \,, m_{l+1})}(u) &\rightarrow
&e^{4\eta (m_l - m_{l+1})} \,, \non \\
B^{(m_n)}(u) &\rightarrow &1 &\,, \qquad &w(u) &\rightarrow  &1  \,, \non \\
z_l(u) &\rightarrow &e^{-2(2n - 2l -1)\eta} &\,, \qquad
&\tilde z_l(u) &\rightarrow &e^{2(2n - 2l -1)\eta} \,. \non
\end{array}
\ee
We conclude that our expression for the transfer matrix eigenvalues
has the correct leading asymptotic behavior for large $u$.

\section{Discussion}

We have obtained an expression for the eigenvalues of the transfer
matrix for the $U_q(B_n)$-invariant $A^{(2)}_{2n}$ open spin chain.
Since this expression passes a number of tests, we are confident
that it is correct. We have not derived the Bethe Ansatz equations
for the model. Instead, we have postulated the Bethe Ansatz equations
to be the ``doubles'' of those for the corresponding closed
chain with periodic boundary conditions. We emphasize that our
result for the eigenvalues of the transfer matrix is {\em not}
simply the double of the corresponding result for the closed chain.
Indeed, our result (\ref{ansatz}), (\ref{z-w}) contains
additional factors that are not present for the closed chain.
Although these new factors have poles, we have seen that the residues
vanish.

It should be straightforward to implement our procedure for the
quantum-algebra invariant open chains associated with the fundamental
representations of $A^{(2)}_{2n - 1}$, $B^{(1)}_{n}$, $C^{(1)}_{n}$
and $D^{(1)}_{n}$. The case $D^{(2)}_{n}$ may be more complicated,
since the $R$ matrix lacks \cite{jimbo} the commutativity property.
Perhaps even the open chains associated with the exceptional affine
algebras can be treated in this way, in view of the fact that the Bethe
Ansatz equations for the closed chains are known \cite{reshetikhin}.

A more satisfactory approach would be the algebraic
Bethe Ansatz, by which the Bethe Ansatz equations can be derived
and also the eigenstates can be constructed. (See, e.g., Ref.
\cite{faddeev/takhtajan}.) Unfortunately, such
an approach is technically difficult to implement for higher-rank
algebras, in particular for open spin chains. (An outline of
the structure of the eigenstates along this line is given in
Appendix B.) In the event that
such a calculation is undertaken, our result for the transfer matrix
eigenvalues should serve as a useful check.

\bigskip

\noindent
{\bf Note added:}

We have evaluated the matrix element (\ref{matrixelement})
for small values of $N$ and $n$ directly, with the help of the symbolic
manipulation program Mathematica. Extrapolating to arbitrary (integer) values
of $N$ and $n$, we find that the pseudovacuum eigenvalue is given by Eq.
(\ref{pseudoeigenvalue}), with $r(u)$ given by
\be
r(u) &=& \frac{\sinh(u)}{\sinh(u-2\eta)}
\frac{\sinh(u-2(2n+1)\eta)}{\sinh(u-4n\eta)}\left[ 1 +
\frac{\sinh(4(n-1)\eta)}{\sinh(2\eta)} \right] \non \\
     &=& w(u) + \sum_{l=1}^{n-1} \left[ z_l(u) + \tilde z_l(u) \right] \non
\ee
where $w(u) \,, z_l(u) \,, \tilde z_l(u)$ are given by Eq. (\ref{z-w}).
This confirms our result (\ref{ansatz}) for the pseudovacuum
eigenvalue. One can now rearrange the argument: assuming that a general
eigenvalue is a ``dressing'' of the pseudovacuum eigenvalue given by
this expression, one can use crossing, the fusion formula, analyticity,
and asymptotic behavior to prove that the Bethe Ansatz equations and the
transfer matrix eigenvalues are precisely the ones given in this paper.

\bigskip

This work was supported in part by the National Science Foundation
under Grant PHY-92 09978.

\appendix

\section{$A^{(2)}_{2n}$ $R$ matrix and its properties}

The $R$ matrix associated with the fundamental representation
of $A_{2n}^{(2)}$ was found by Bazhanov \cite{bazhanov} and
Jimbo \cite{jimbo}. We follow the latter reference; however, we use the
variables $u$ and $\eta$ instead of $x$ and $k$, respectively, which
are related as follows:
\be
x = e^u \,, \qquad \qquad k = e^{2 \eta} \,.
\ee
The $R$ matrix is given by \footnote{This expression for the $R$ matrix
differs from the one given in Ref. \cite{jimbo} by the overall factor
$2 e^{u + (2n+3)\eta}$.}
\noi
\be
R(u)&=&
c(u)\sum_{\alpha \neq \alpha'}E_{\alpha \alpha} \otimes E_{\alpha \alpha}
+ b(u) \sum_{\alpha \neq\beta,\beta'} E_{\alpha \alpha} \otimes
     E_{\beta \beta} \\
& + & (e(u) \sum_{\alpha < \beta, \alpha \neq \beta'}
+\bar e(u) \sum_{\alpha > \beta,\alpha\neq \beta'})E_{\alpha \beta}
\otimes E_{\beta \alpha}
+\sum_{\alpha \,, \beta} a_{\alpha \beta}(u) E_{\alpha \beta}
\otimes E_{\alpha' \beta'} \non
\ee
with
\be
c(u)&=&2 \sinh(\uh-2\eta)\ \cosh(\uh - (2n+1)\eta) \non \\
b(u)&=&2 \sinh(\uh)\ \cosh(\uh - (2n+1)\eta) \label{factors} \\
e(u)&=&-2 e^{-\uh} \sinh (2\eta)\ \cosh(\uh -(2n+1)\eta) \non \\
\bar e(u)&=&e^u e(u) \non
\ee
\be
a_{\alpha \beta}(u)=\left\{ \begin{array}{ll}
\sinh(u+(-2n+1)\eta)+\sinh((2n-1)\eta) & \alpha =\beta, \alpha \neq \alpha'\\
\non\\
\sinh(u-(2n+1)\eta)+\sinh(2n+1)\eta+\\+\sinh((2n-1)\eta)-\sinh((2n+3)\eta) &
\alpha=\beta, \alpha=\alpha'\\
\non\\
-2 e^{((2n+1)+2(\bar{\alpha}-\bar{\beta}))\eta}e^{-\uh}\sinh\uh \sinh 2\eta &
\alpha <\beta, \alpha\neq \beta'\\
\non\\
2 e^{-u+(2(2n+1)-2\beta+2)\eta}\sinh(2n+3-2 \beta)\eta \sinh 2\eta-\\-
2\sinh 2\eta \cosh(2(2n+2)-2 \beta)\eta e^{((2n+3)-2 \beta)\eta} &
\alpha <\beta, \alpha=\beta'\\
\non\\
2 e^{(-(2n+1)+2(\bar{\alpha}-\bar{\beta}))\eta} e^\uh \sinh \uh \sinh 2\eta &
\alpha>\beta, \alpha\neq \beta'\\
\non\\
2 e^{u-2 \beta\eta}\sinh ((2n+1)-2 \beta)\eta \sinh 2\eta-\\-2\sinh 2\eta
\cosh(2\beta\eta) e^{((2n+1)-2\beta)\eta} & \alpha>\beta, \alpha=\beta'
\end{array} \right.
\label{Ra}
\ee
where
\be
\bar{\alpha}=\left\{ \begin{array}{ll}
\alpha+\half & 1\le\alpha<n+1\\
\alpha & \alpha=n+1\\
\alpha-\half & n+1<\alpha\le 2n+1
\end{array} \right.
\ee
\be
\alpha,\beta &=& 1 \,, \cdots \,, 2n+1 \non \\
\alpha' &=& 2n+2-\alpha
\ee
and the $E_{\alpha \beta}$ are elementary matrices.
Evidently, the $R$ matrix acts on the tensor product space
$C^{2n+1} \otimes C^{2n+1}$.

In addition to obeying the Yang-Baxter equation (\ref{yang-baxter}),
this $R$ matrix satisfies the following important properties:

\noi{$PT\ $ {\it symmetry}}
\be
{\cal P}_{12}\ R_{12}(u)\ {\cal P}_{12} \equiv R_{21}(u)
= R_{12}(u)^{t_1 t_2} \,;
\label{pt}
\ee

\noi{\it unitarity }
\be
R_{12}(u)\ R_{21}(-u)&=&\zeta(u) \,, \label{unitarity}
\ee
where $\zeta (u)$ is given by
\be
 \zeta(u)=-4\sinh(\uh -2\eta) \cosh(\uh -(2n+1)\eta)
            \sinh(\uh +2\eta) \cosh(\uh +(2n+1)\eta)\,;
\ee

\noi{\it crossing symmetry}
\be
R_{12}(u)=V_1\ R_{12}(-u-\rho)^{t_2}\ V_1
= V_2^{t_2}\ R_{12}(-u-\rho)^{t_1}\ V_2^{t_2} \,,
\label{cross}
\ee
where $\rho=-i \pi -2(2n+1)\eta$ and $V^2 = 1$;

\noi{\it regularity}
\be
R(0)= -\zeta(0)^\half {\cal P}  \,,
\ee
where ${\cal P}$ is the permutation operator
\be
{\cal P}=\sum_{\alpha, \beta} E_{\alpha \beta}\otimes E_{\beta \alpha}
\,;
\ee

\noi{\it commutativity}
\be
\left[ \check{R}(u) \,, \check{R}(v) \right] = 0,
\hspace{2cm} \check{R} = \cal{P} R \,; \label{Rcheck}
\ee

\noi{\it periodicity}
\be
R(u + 2\pi i) = R(u)  \,.
\ee

In order to determine the crossing matrix $V$, we begin by
making the following ansatz:
\be
V =\sum_\alpha  v_{\alpha \alpha'} E_{\alpha \alpha'}
\ee
where $\alpha' = 2n + 2 - \alpha$. The constraint $V^2 = 1$ implies that
\be
v_{\alpha \alpha'} v_{\alpha' \alpha}=1 \,.
\ee
(Note that we do {\em not} sum over repeated indices.)
Inserting this ansatz into the crossing equation (\ref{cross}) leads
to the the following relations:
\be
\left( v_{\alpha\alpha'} \right)^2 &=&
\frac{a_{\alpha\alpha'}(u)}{a_{\alpha'\alpha}(-u-\rho)} \,,
\qquad \alpha \ne \alpha' \label{compV1} \\
e(u) &=& v_{\alpha'\alpha} v_{\beta\beta'} a_{\alpha\beta}(-u-\rho) \,,
\qquad \alpha > \beta \,, \quad \beta \ne \alpha' \,, \label{compV2}
\ee
with $a_{\alpha\beta}(u)$ and $e(u)$ defined in Eqs. (\ref{Ra})
and (\ref{factors}), respectively. The first of these relations
determines $v_{\alpha\alpha'}$ up to signs; while the second relation
fixes the signs, up to an overall sign which we specify by the condition
$v_{\alpha\alpha'}=1$ for $\alpha = \alpha'$.
This leads to the result
\be
V= E_{\alpha \alpha}
-\sum_{\alpha<\alpha'} e^{(-(2n+1)+2 \alpha)\eta} E_{\alpha \alpha'}
-\sum_{\alpha>\alpha'} e^{(2n+1-2 \alpha')\eta} E_{\alpha \alpha'} \,.
\label{V}
\ee
Correspondingly, $M=V^t\ V$ is given by the $(2n+1) \times (2n+1)$
diagonal matrix
\be
M = diag \left( e^{2(2n-1)\eta} \,, e^{2(2n-3)\eta} \,, \cdots \,,
e^{2\eta} \,, 1 \,, e^{-2\eta} \,, \cdots \,, e^{-2(2n-3)\eta} \,,
e^{-2(2n-1)\eta} \right) \label{M} \,.
\ee

Finally, we note that the leading asymptotic behavior of
$R(u)$ for large $u$ is given by a lower-triangular matrix
\be
R_{a k}(u) \stackrel{u \rightarrow \infty}{\sim}
\half e^{u-(2n+1)\eta}
\left( \begin{array}{lcccccr}
e^{-2\eta H_{1 (k)}} &       &         &      &        &     & \\
      &      \ddots    &         &      &        0     & \\
      &        &  e^{-2\eta H_{n (k)}} &      &        &     & \\
      &        &       &         1      &        &     & \\
      &        &       &         & e^{2\eta H_{n (k)}} &     & \\
      &      \star     &         &      &      \ddots  & \\
      &        &       &         &      &        & e^{2\eta H_{1 (k)}}
\end{array}\right) \label{Rasymptotic} \,,
\ee
where $H_{1 (k)} \,, \cdots \,, H_{n (k)}$ are the $(2n+1) \times (2n+1)$
diagonal matrices given in Eq. (\ref{hl}), and $1$ is the
$(2n+1) \times (2n+1)$ unit matrix.

\section{Structure of eigenstates}

We outline here the structure of the eigenstates of the transfer matrix.
In particular, we explain the formula (\ref{r}) for the eigenvalues of
the Cartan generators.

In the algebraic Bethe Ansatz approach, one identifies
certain operators with which one constructs the eigenstates.
A detailed investigation along this line for the $U_q(B_n)$-invariant
$A^{(2)}_{2n}$ spin chain is clearly outside the scope of this paper.
Nevertheless, certain features can already be anticipated.
We expect to have tensor operators $B^{(l)}(u)$ which are related to
the lowering operators $E_{- \alpha_l}$ of $U_q(B_n)$,
\be
B^{(l)}(u) \sim E_{- \alpha_l} \,, \qquad l = 1\,, \cdots \,, n
\ee
where  $\alpha_l$ are the simple roots of $B_n$,
\be
\alpha_1 &=& (1 \,, -1 \,, 0 \,, \cdots \,, 0) \non \\
\alpha_2 &=& (0 \,, 1 \,, -1 \,, 0 \,, \cdots \,, 0) \non \\
         &\vdots&                                    \\
\alpha_{n-1} &=& (0 \,, \cdots \,, 0 \,, 1 \,, -1) \non \\
\alpha_{n} &=& (0 \,, \cdots \,, 0 \,, 1)  \,. \non
\ee
In particular, $B^{(l)}(u)$ and $E_{- \alpha_l}$ have the same
commutation relations with the Cartan generators:
\be
\left[ H_1 \,, B^{(1)}(u) \right] &= - B^{(1)}(u) \,, \qquad
\left[ H_2 \,, B^{(1)}(u) \right] &=   B^{(1)}(u) \,, \non \\
\left[ H_2 \,, B^{(2)}(u) \right] &= - B^{(2)}(u) \,, \qquad
\left[ H_3 \,, B^{(2)}(u) \right] &=   B^{(2)}(u) \,, \non \\
& \vdots \label{comm}
\ee

The Bethe states (eigenstates of the transfer matrix) are constructed
by acting with the operators $B^{(l)}(u)$ on the pseudovacuum state,
\be
| \Lambda^{(m_1 \,, \cdots \,, m_n)} > &=&
B^{(1)}(u^{(1)}_1) \cdots B^{(1)}(u^{(1)}_{m_1})\
B^{(2)}(u^{(2)}_1) \cdots B^{(2)}(u^{(2)}_{m_2})\
\cdots \non \\
& & \times
B^{(n)}(u^{(n)}_1) \cdots B^{(n)}(u^{(n)}_{m_n})\
| \Lambda^{(0 \,, \cdots \,, 0)} > \,, \label{bethe}
\ee
where $u^{(l)}_j$ are solutions of the Bethe Ansatz Eqs. (\ref{ba}).
The pseudovacuum state (\ref{up}) evidently satisfies
\be
H_1\ | \Lambda^{(0\,, \cdots \,, 0)} >
&=& N | \Lambda^{(0 \,, \cdots \,, 0)} > \,, \non \\
H_l\ | \Lambda^{(0 \,, \cdots \,, 0)} >
&=& 0 \,, \qquad l = 2 \,, \cdots \,, n \,.
\ee
Using the commutation relations (\ref{comm}), one can readily
compute the eigenvalues of the Cartan generators on the
Bethe states (\ref{bethe}), and obtain the result given in
Eq. (\ref{r}).

\vfill\eject

\end{document}